\documentclass[preprint,12pt]{aastex}

\usepackage{epsf}

\newcommand{\bT}{\mbox{$\bf T$}}
\newcommand{\bu}{\mbox{\boldmath$u$}}

\begin{document}

\title{Saturation of the corotation resonance in a gaseous disk}
\shorttitle{Saturation of the corotation resonance}
\author{G. I. Ogilvie\altaffilmark{1} and S. H. Lubow\altaffilmark{2}}
\altaffiltext{1}{Institute of Astronomy, University of Cambridge,
  Madingley Road, Cambridge CB3 0HA, UK; {\tt gogilvie@ast.cam.ac.uk}}
\altaffiltext{2}{Space Telescope Science Institute,
  3700 San Martin Drive, Baltimore, MD 21218; {\tt lubow@stsci.edu}}
\shortauthors{Ogilvie \& Lubow}

\begin{abstract}
  We determine the torque exerted in a steady state by an external
  potential on a three-dimensional gaseous disk at a non-coorbital
  corotation resonance.  Our model accounts for the feedback of the
  torque on the surface density and vorticity in the corotation
  region, and assumes that the disk has a barotropic equation of state
  and a nonzero effective viscosity.  The ratio of the torque to the
  value given by the formula of Goldreich \& Tremaine depends
  essentially on a single dimensionless parameter, which quantifies
  the extent to which the resonance is saturated.  We discuss the
  implications for the eccentricity evolution of young planets.
\end{abstract}


\section{Introduction}

The forcing of a gaseous disk by a gravitational perturber at a
resonance can result in a strong response and an interchange of energy
and angular momentum between the perturber and the disk.  A corotation
resonance arises where the pattern speed of the forcing matches the
local angular speed of disk material.  This occurs in the context of
galaxies where a stellar bar or spiral arms force motions in the
interstellar medium (Binney \& Tremaine 1987).  Corotation resonances
also arise when satellites orbit within a disk, as occurs with young
planets in protoplanetary disks, or moons in planetary ring systems
(Goldreich \& Tremaine 1979, 1980; hereafter GT79 and GT80
respectively).

For planets, two types of corotation resonance need to be
distinguished.  A coorbital corotation resonance occurs where the
orbital period of the planet matches the orbital period of disk
material. For a planet with a circular orbit, this is the only type of
corotation resonance that can arise.  The second type of corotation
resonance is non-coorbital and occurs if the planet executes an
eccentric orbit.  The forcing due to a planet with an eccentric orbit
can be decomposed into a series of rotating components of various
strengths and pattern speeds (GT80).  Bar or spiral galaxies, on the
other hand, give rise to non-coorbital resonances.

The analyses of the coorbital and non-coorbital resonances are
different.  In the coorbital case, which has recently been considered
by Masset (2001, 2002) and Balmforth \& Korycansky (2001), the forcing
is stronger and involves multiple Fourier components.  There is more
likely to be a reduction in density caused by the tendency of the
planet to open a gap in the disk.  In this paper, however, we are
concerned with non-coorbital corotation resonances, which are critical
in determining the eccentricity evolution of planets resulting from
planet--disk interactions (GT80), and are therefore likely to be
important in explaining the eccentricities of many of the observed
extrasolar planets.  For a disk in which a gap is opened by a planet
having a small orbital eccentricity, corotation resonances tend to
damp the planet's eccentricity, while Lindblad resonances cause it to
grow.  The effects of each type of resonance are strong, but are
nearly equal in magnitude, to the extent that they nearly cancel.  The
balance is slightly in favor of eccentricity damping, if the
corotation resonances operate at maximal efficiency (i.e., are
unsaturated).  The final outcome of eccentricity evolution depends on
the details.

The disturbances of the disk caused by a corotation resonance remain
localized to the corotation region (GT79). They are unable to
propagate away from the resonance, as occurs in the case of Lindblad
resonances.  Instead, they act back locally on the disk and may change
its density and vorticity in such a manner as to reduce (saturate) the
corotation torque (e.g., Ward 1991).  However, the effective turbulent
viscosity of the disk could lessen the saturation by limiting the
back-reaction on the disk (Ward 1992).  Full saturation occurs when
the torque is reduced to zero, but even a small degree of saturation
(5\%) could change the sense of eccentricity evolution in the GT80
model from decay to growth (Goldreich \& Sari 2002).

In this paper, we develop a detailed model for the saturation of a
non-coorbital corotation resonance in a viscous accretion disk. We
determine how the strength of the corotation torque varies with the
strength of the tidal forcing and the effective viscosity of the disk.

\section{Physical picture}

The interaction between the perturber and the disk in the corotation
region can be considered to involve two distinct aspects.  In one
aspect, the tidal potential excites an evanescent disturbance in the
disk, i.e., a disturbance that decays exponentially with distance from
the resonance over the corotation region.  Associated with this
disturbance are evanescent angular momentum fluxes that are oppositely
directed on the two sides of the resonance.  In the linear, inviscid
theory of GT79, there is a jump in the flux at the resonance that
corresponds to a torque being injected into the disk exactly at the
corotation radius.  The torque depends on the radial derivative of the
ratio of the surface density to the vorticity at corotation.  In the
second aspect, the disk responds to the torque by changing the
distributions of surface density and angular velocity over the
corotation region (Lubow 1990).  These changes are such as to cause a
reduction in the torque, and, ultimately, it is expected that the
torque will be eliminated.  This picture is time-dependent because the
torque injection continues until full saturation occurs.

In the case of an accretion disk, full saturation may not occur owing
to the effects of turbulent viscosity in limiting the back-reaction.
We therefore aim to find steady-state solutions in which the torque on
the disk is nonzero.  We also allow simultaneously for both aspects
described above, by formulating a nonlinear problem in which the
feedback is included self-consistently.  In the process of the
analysis, we will determine the size of the region over which the
torque is injected into the disk.

\section{Analysis of the corotation region}

\subsection{Basic equations}

We consider a three-dimensional gaseous disk having a barotropic
equation of state, so that the pressure $p$ depends only on the
density $\rho$.  We assume that the forcing potential and the
properties of the unperturbed disk vary smoothly in radius across the
corotation region.  These assumptions may need to be reconsidered if
the resonance is very close to the planet (i.e., at a distance
comparable to the vertical scale-height $H$ of the disk), or close to
a sharp disk edge (where the density varies on a radial scale
comparable to $H$).  However, the torque cut-off (GT79) limits the
effectiveness of resonances that are closer than approximately $H$,
and a gap in the disk about the planet's orbit may eliminate such
close resonances in any case.  We also consider the planet to execute
a fixed orbit, and ignore any effects of migration.

The disk is governed by the equation of mass conservation,
\begin{equation}
  (\partial_t+\bu\cdot\nabla)\rho=-\rho\nabla\cdot\bu,
\end{equation}
and the equation of motion,
\begin{equation}
  (\partial_t+\bu\cdot\nabla)\bu=-\nabla(\Phi+h)+
  {{1}\over{\rho}}\nabla\cdot\bT.
\end{equation}
Here $\bu$ is the velocity, $\Phi$ the gravitational potential,
$h=\int\rho^{-1}\,dp$ the enthalpy, and $\bT$ the stress tensor, which
represents a viscous or turbulent stress.  For a Navier--Stokes model
we have
\begin{equation}
  \bT=\mu\left[\nabla\bu+(\nabla\bu)^{\rm T}\right]+
  (\mu_{\rm b}-{\textstyle{{2}\over{3}}}\mu)(\nabla\cdot\bu){\bf1},
\end{equation}
where $\mu$ is the shear viscosity and $\mu_{\rm b}$ the bulk
viscosity.  We assume that, as is conventional in accretion disk
theory, the vertically integrated shear viscosity $\nu\Sigma$ may be
expressed as a function of the radius $r$ and the surface density
$\Sigma$.  The logarithmic derivative
$D_\mu=\partial\ln(\nu\Sigma)/\partial\ln\Sigma$ will feature in the
analysis below.

The unperturbed disk is a steady, axisymmetric solution
$\{\rho^{\rm(u)},\bu^{\rm(u)},h^{\rm(u)},\bT^{\rm(u)}\}$ of these
equations in the presence of the steady, axisymmetric potential
$\Phi^{\rm(u)}$ of the central mass.  Let the potential be perturbed
by the addition of a nonaxisymmetric external potential $\Phi'$ that
rotates rigidly with angular pattern speed $\Omega_{\rm p}$.  The
solution in the presence of the perturbed potential
$\Phi^{\rm(p)}=\Phi^{\rm(u)}+\Phi'$ is
$\{\rho^{\rm(p)},\bu^{\rm(p)},h^{\rm(p)},\bT^{\rm(p)}\}$, and may be
assumed to be steady in a frame of reference that rotates at
$\Omega_{\rm p}$.

We write the nonlinear equations for the Eulerian perturbations
$\rho'=\rho^{\rm(p)}-\rho^{\rm(u)}$, etc., in cylindrical polar
coordinates $(r,\phi,z)$ in the rotating frame, using $(u,v,w)$ to
denote the cylindrical velocity components, and omitting the
superscript (u):
\begin{equation}
  D\rho'+(u'\partial_r+w'\partial_z)\rho=
  -\rho'\left[{{1}\over{r}}\partial_r(ru)+\partial_zw\right]-
  (\rho+\rho')\left[{{1}\over{r}}\partial_r(ru')+
  {{1}\over{r}}\partial_\phi v'+\partial_zw'\right],
  \label{rho'}
\end{equation}
\begin{equation}
  Du'+(u'\partial_r+w'\partial_z)u-2\Omega v'-{{{v'}^2}\over{r}}=
  -\partial_r(\Phi'+h')+\cdots,
  \label{u'}
\end{equation}
\begin{equation}
  {{1}\over{r}}D(rv')+2Bu'=-{{1}\over{r}}\partial_\phi(\Phi'+h')+\cdots,
  \label{v'}
\end{equation}
\begin{equation}
  Dw'+(u'\partial_r+w'\partial_z)w=-\partial_z(\Phi'+h')+\cdots.
  \label{w'}
\end{equation}
Here
\begin{equation}
  D=(u+u')\partial_r+\left(\Omega-\Omega_{\rm p}+
  {{v'}\over{r}}\right)\partial_\phi+(w+w')\partial_z,
\end{equation}
and the dots denote viscous terms that we do not write out owing to
their complexity.  $\Omega=v/r$ is the unperturbed angular velocity in the
inertial frame, which depends only on $r$ in a barotropic disk, and
\begin{equation}
  2B={{1}\over{r}}{{d}\over{dr}}(r^2\Omega)
\end{equation}
is the vertical vorticity, or twice Oort's second parameter.  The
epicyclic frequency $\kappa$ is given by $\kappa^2=4\Omega B$.
Also
\begin{equation}
  h'={{v_{\rm s}^2}\over{\rho}}\rho'+O({\rho'}^2),
  \label{h'}
\end{equation}
where $v_{\rm s}^2=dp/d\rho$ is the square of the sound speed.

\subsection{Expansion about corotation}

\label{Expansion about corotation}

Let $r_{\rm c}$ be the corotation radius, defined by $\Omega(r_{\rm
  c})=\Omega_{\rm p}$.  There are three characteristic length-scales
that are potentially relevant for determining the radial extent of the
corotation region,
\begin{equation}
  \delta_c={{c}\over{\kappa}},\qquad
  \delta_\nu=\left({{\nu}\over{-m\,d\Omega/dr}}\right)^{1/3},\qquad
  \delta_\Psi=\left({{\Psi}\over{-\kappa^2\,d\ln\Omega/d\ln r}}\right)^{1/2}.
  \label{delta}
\end{equation}
Here $c$ is an appropriate average of the sound speed $v_{\rm s}$,
while $m$ and $\Psi$ are the azimuthal wavenumber and amplitude of the
forcing potential $\Phi'$.  We assume that $d\Omega/dr<0$.  In a
linear, inviscid theory (GT79) the scale is set by the evanescent
wavelength $\delta_c$ of the two-dimensional mode (density wave) near
corotation.  However, the solution involves a cusp, which may be
expected to be smoothed over a characteristic length $\delta_\nu$ in the
presence of an effective viscosity.  On the other hand, a ballistic
treatment of the corotation region suggests that an annulus of
librating trajectories is formed, of half-width $2\sqrt{2}\delta_\Psi$
(Goldreich \& Tremaine 1981).

In order to allow all these effects to compete, we adopt scalings such
that these three lengths are formally of the same order of magnitude.
(Subsequently, we will take limits in which one or other of the
quantities $\delta$ can be distinguished as the relevant scale.)  The
characteristic width of the corotation region is then $H$, the
semi-thickness of the disk, as was also found by Lubow (1990).  This
allows us to resolve the radial structure of the region in the same
way as for the vertical structure, by introducing scaled coordinates
\begin{equation}
  \xi={{r-r_{\rm c}}\over{\epsilon}},\qquad
  \zeta={{z}\over{\epsilon}},
\end{equation}
where $\epsilon\ll1$ is a typical value of the angular semi-thickness
$H/r$ of the disk.  In units such that $r_{\rm c}=1$, the corotation
region is found where $\xi$ and $\zeta$ are of order unity.

The central potential may be expanded in the corotation region in a
double Taylor series,
\begin{equation}
  \Phi=(\Phi_{00}+\epsilon\Phi_{01}\xi+\cdots)+
  \epsilon^2{\textstyle{{1}\over{2}}}\zeta^2
  (\Phi_{20}+\epsilon\Phi_{21}\xi+\cdots)+\cdots.
\end{equation}
The perturbing potential may be expanded similarly, but is assumed
formally to be smaller by a factor $O(\epsilon^2)$:
\begin{equation}
  \Phi'/\epsilon^2=[\Phi'_{00}(\phi)+
  \epsilon\Phi'_{01}(\phi)\xi+\cdots]+
  \epsilon^2{\textstyle{{1}\over{2}}}\zeta^2
  [\Phi'_{20}(\phi)+\epsilon\Phi'_{21}(\phi)\xi+\cdots]+\cdots.
\end{equation}
This assumes that both potentials are symmetric about $z=0$ and vary
smoothly in $r$ in the neighborhood of $r_{\rm c}$.  The scaling of
$\Phi'$ will turn out to provide a critical level of nonlinearity in
the solution.

Properties of the unperturbed disk generally have non-trivial
structure in $\zeta$ but may be expanded in Taylor series in $r$,
\begin{eqnarray}
  \rho&=&\rho_0(\zeta)+\epsilon\rho_1(\zeta)\xi+\cdots,\nonumber\\
  u&=&\epsilon^3\left[u_0(\zeta)+\cdots\right],\nonumber\\
  w&=&\epsilon^4\left[w_0(\zeta)+\cdots\right],\nonumber\\
  \Omega&=&\Omega_0+\epsilon\Omega_1\xi+\cdots,\nonumber\\
  B&=&B_0+\epsilon B_1\xi+\cdots,\nonumber\\
  v_{\rm s}&=&\epsilon\left[v_{{\rm s}0}(\zeta)+\cdots\right],\nonumber\\
  \mu&=&\epsilon^3\left[\mu_0(\zeta)+\cdots\right],\nonumber\\
  \mu_{\rm b}&=&\epsilon^3\left[\mu_{{\rm b}0}(\zeta)+\cdots\right].
\end{eqnarray}
Note that $\Omega_0=\Omega_{\rm p}$ by definition of the corotation
radius.  The scaling of $\mu$ implies that the dimensionless viscosity
parameter, $\alpha=\mu\Omega/p$, is formally $O(\epsilon)$, and will
turn out to provide a critical level of dissipation in the solution.
The scalings of the accretion flow, $u$ and $w$, follow from this
assumption.

Perturbed variables generally have non-trivial structure in $\xi$ and
$\zeta$.  We have found the required scalings to be
\begin{eqnarray}
  \rho'&=&\epsilon\left[\rho'_0(\xi,\phi,\zeta)+\cdots\right],\nonumber\\
  u'&=&\epsilon^2\left[u'_0(\phi)+\epsilon u'_1(\xi,\phi,\zeta)+
  \cdots\right],\nonumber\\
  v'&=&\epsilon^2\left[v'_0(\xi,\phi)+\epsilon v'_1(\xi,\phi,\zeta)+
  \cdots\right],\nonumber\\
  w'&=&\epsilon^3\left[w'_0(\xi,\phi,\zeta)+\cdots\right],\nonumber\\
  h'&=&\epsilon^3\left[h'_0(\xi,\phi)+\epsilon h'_1(\xi,\phi,\zeta)+
  \cdots\right],\nonumber\\
  \mu'&=&\epsilon^4\left[\mu'_0(\xi,\phi,\zeta)+\cdots\right],\nonumber\\
  \mu'_{\rm b}&=&\epsilon^4\left[\mu'_{{\rm b}0}(\xi,\phi,\zeta)+
  \cdots\right].
\end{eqnarray}

Equations (\ref{rho'}), (\ref{u'}) and (\ref{v'}) at leading order,
$O(\epsilon^2)$, then imply
\begin{equation}
  (u'_0\partial_\xi+\Omega_1\xi\partial_\phi)\rho'_0+u'_0\rho_1+
  w'_0\partial_\zeta\rho_0=
  -\rho_0\left({{u'_0}\over{r_{\rm c}}}+\partial_\xi u'_1+
  {{1}\over{r_{\rm c}}}\partial_\phi v'_0+\partial_\zeta w'_0\right),
  \label{eq1}
\end{equation}
\begin{equation}
  -2\Omega_0v'_0=-\Phi'_{01}-\partial_\xi h'_0,
  \label{eq2}
\end{equation}
\begin{equation}
  2B_0u'_0=-{{1}\over{r_{\rm c}}}\partial_\phi\Phi'_{00}.
  \label{eq3}
\end{equation}
Also required is equation (\ref{v'}) at the next order, $O(\epsilon^3)$,
\begin{equation}
  (u'_0\partial_\xi+\Omega_1\xi\partial_\phi)v'_0+2B_0u'_1+2B_1\xi u'_0=
  -{{\xi}\over{r_{\rm c}}}\left(\partial_\phi\Phi'_{01}-
  {{1}\over{r_{\rm c}}}\partial_\phi\Phi'_{00}\right)
  -{{1}\over{r_{\rm c}}}\partial_\phi h'_0+
  {{\mu_0}\over{\rho_0}}\partial_\xi^2v'_0+
  {{r_{\rm c}\Omega_1}\over{\rho_0}}\partial_\xi\mu'_0.
  \label{eq4}
\end{equation}
The last two terms are the only viscous terms to enter our analysis.
Finally, equation (\ref{w'}) at leading order, $O(\epsilon^3)$,
requires only a hydrostatic balance,
\begin{equation}
  0=-\Phi'_{20}\zeta-\partial_\zeta h'_1.
  \label{eq5}
\end{equation}
The enthalpy and density perturbations are related by equation
(\ref{h'}) at leading order, $O(\epsilon^3)$,
\begin{equation}
  h'_0={{v_{{\rm s}0}^2}\over{\rho_0}}\rho'_0.
  \label{eq6}
\end{equation}

\subsection{Derivation of the governing equation}

We eliminate $\rho'_0$, $u'_1$, $v'_0$, and $w'_0$ (by integrating
eq.~[\ref{eq1}] with respect to $\zeta$ over the full vertical extent
of the disk) to obtain a single equation for $h'_0$,
\begin{equation}
  (u'_0\partial_\xi+\Omega_1\xi\partial_\phi)
  (\kappa_0^2I_3-I_1\partial_\xi^2)h'_0+
  (I_4\partial_\xi^2+2r_{\rm c}\Omega_0\Omega_1I_5)\partial_\xi^2h'_0
  ={{2\Omega_0}\over{r_{\rm c}}}(\partial_\phi\Phi'_{00})
  \left(I_2-{{B_1}\over{B_0}}I_1\right),
\end{equation}
involving the integrals
\begin{equation}
  I_1=\int\rho_0\,d\zeta,\qquad
  I_2=\int\rho_1\,d\zeta,\qquad
  I_3=\int{{\rho_0}\over{v_{{\rm s}0}^2}}\,d\zeta,\qquad
  I_4=\int\mu_0\,d\zeta,\qquad
  I_5={{1}\over{h'_0}}\int\mu'_0\,d\zeta.
\end{equation}
Here $u'_0$ is given directly by equation (\ref{eq3}).  We recognize
$I_1=\Sigma_0$ and $I_2=\Sigma_1$ as coefficients in the Taylor
expansion of the unperturbed surface density $\Sigma(r)$.  Defining
the mean sound speed $c$, the mean kinematic viscosity $\nu$, and the
logarithmic viscosity derivative $D_\mu$ in physical terms by
\begin{equation}
  {{\Sigma}\over{c^2}}=\int{{\rho}\over{v_{\rm s}^2}}\,dz,\qquad
  \nu\Sigma=\int\mu\,dz,\qquad
  D_\mu={{\partial\ln(\nu\Sigma)}\over{\partial\ln\Sigma}},
\end{equation}
we find $I_3=\Sigma_0/c_0^2$, $I_4=\nu_0\Sigma_0$, and
$I_5=D_\mu\nu_0\Sigma_0/c_0^2$.

When the $\epsilon$-scalings and subscripts are omitted, the final
equation in physical terms reads
\begin{equation}
  \left(u'\partial_\xi+{{d\Omega}\over{dr}}\xi\partial_\phi\right)
  \left({{\kappa^2}\over{c^2}}-\partial_\xi^2\right)h'+
  \nu\left(\partial_\xi^2+2r\Omega{{d\Omega}\over{dr}}
  {{D_\mu}\over{c^2}}\right)\partial_\xi^2h'=
  {{2\Omega}\over{r}}(\partial_\phi\Phi')
  {{d}\over{dr}}\ln\left({{\Sigma}\over{B}}\right),
  \label{final}
\end{equation}
where
\begin{equation}
  u'=-{{1}\over{2rB}}\partial_\phi\Phi'.
\end{equation}
Here, all quantities except $\xi$, $\Phi'(\phi)$, $u'(\phi)$, and
$h'(\xi,\phi)$ are to be understood as constants evaluated at
$r=r_{\rm c}$.

Equation (\ref{final}) describes, in a reduced but systematically
obtained manner, how the enthalpy perturbation is forced by the
external potential in the corotation region.  If the advective term
involving $u'$ were omitted, this would be a linear inhomogeneous
equation, and the forcing would be proportional to the gradient of the
vortensity $\Sigma/B$.  The $u'$ term derives from a product of
perturbed quantities and is in this sense a nonlinear effect.  An
alternative way to write equation (\ref{final}) is
\begin{equation}
  {{d\Omega}\over{dr}}\xi\partial_\phi
  \left({{\kappa^2}\over{c^2}}-\partial_\xi^2\right)h'+
  \nu\left(\partial_\xi^2+2r\Omega{{d\Omega}\over{dr}}
  {{D_\mu}\over{c^2}}\right)\partial_\xi^2h'=
  {{2\Omega}\over{r}}(\partial_\phi\Phi')
  \left[{{d}\over{dr}}\ln\left({{\Sigma}\over{B}}\right)+
  \partial_\xi\left({{\Sigma'}\over{\Sigma}}-{{B'}\over{B}}\right)\right],
\end{equation}
where $\Sigma'=\Sigma h'/c^2$ and $2B'=\partial_\xi v'$ are the
leading-order perturbations of the surface density and the vertical
vorticity.  This shows that the nonlinear effect can be understood
either as an additional advection of the perturbation or as the
feedback of the perturbation on the vortensity gradient of the disk.

\subsection{Reduction to dimensionless form}

\label{Reduction to dimensionless form}

We assume that the external potential contains a single Fourier
component, $\Phi'=\Psi\cos m\phi$, and rewrite the governing equation
(\ref{final}) in a dimensionless form by means of the transformations
\begin{equation}
  \xi=x\,{{c}\over{\kappa}},\qquad
  \phi={{\theta}\over{m}},\qquad
  h'(\xi,\phi)=f(x,\theta)\,{{c^3}\over{\kappa}}
  {{d}\over{dr}}\ln\left({{\Sigma}\over{B}}\right).
\end{equation}
Thus
\begin{equation}
  (-a\sin\theta\,\partial_x+
  x\partial_{\theta})
  (1-\partial_x^2)f-\tilde\nu(\partial_x^2-b)
  \partial_x^2f=a\sin\theta,
  \label{dimensionless}
\end{equation}
where
\begin{eqnarray}
  a&=&2\left(-{{d\ln r}\over{d\ln\Omega}}\right){{\Psi}\over{c^2}},\nonumber\\
  b&=&-2r\Omega{{d\Omega}\over{dr}}{{D_\mu}\over{\kappa^2}},\nonumber\\
  \tilde\nu&=&\left({{\nu}\over{-m\,d\Omega/dr}}\right)
  {{\kappa^3}\over{c^3}}.
\end{eqnarray}
The problem therefore depends on only three dimensionless parameters:
the forcing amplitude $a$, the viscosity derivative $b$, and the
dimensionless viscosity $\tilde\nu$.  For a Keplerian disk,
$b=3D_\mu$.  In terms of the characteristic length-scales defined in
equation (\ref{delta}), $\tilde\nu=(\delta_\nu/\delta_c)^3$.

Equation (\ref{dimensionless}) is to be solved on
$-\infty<x<\infty$, $0<\theta<2\pi$ subject to a periodic boundary
condition in $\theta$, which implies that the physical solution has
the same periodicity in $\phi$ as the forcing potential.  The solution
$f$ should be bounded as $|x|\to\infty$ in order to be a valid
localized solution in the corotation region.

We note immediately that one non-localized solution of equation
(\ref{dimensionless}) is $f=-x+{\rm constant}$.  This represents an
axisymmetric perturbation that exactly cancels the vortensity gradient
of the unperturbed disk, so nullifying the corotation resonance.  Such
a solution should be excluded, as it amounts to redefining the
original problem.

\subsection{Solution by Fourier analysis}

Equation (\ref{dimensionless}) contains multiple $x$-derivatives but
involves only one power of $x$ explicitly.  Like Airy's equation, it
is amenable to Fourier analysis in $x$.  Let
\begin{equation}
  F(k,\theta)=\int_{-\infty}^\infty f(x,\theta)\,
  e^{-ikx}\,dx
\end{equation}
be the Fourier transform of $f$, which exists provided that $f$ decays
sufficiently rapidly as $|x|\to\infty$.  (In fact, by allowing for $F$
to be a generalized function, we may permit $f$ to grow at most
algebraically in $x$.)  The equation transforms to
\begin{equation}
  \left[-i\partial_k\partial_\theta+ika\sin\theta+
  \tilde\nu k^2\left({{b+k^2}\over{1+k^2}}\right)\right]G=
  2i\sin\theta\,\delta(k),
  \label{k-phi}
\end{equation}
where
\begin{equation}
  G(k,\theta)={{1}\over{i\pi a}}(1+k^2)F(k,\theta).
\end{equation}
Equation (\ref{k-phi}) could be solved numerically as a partial
differential equation, but is amenable to further Fourier analysis in
$\theta$.  Let
\begin{equation}
  G(k,\theta)=\sum_{n=-\infty}^\infty G_n(k)\,e^{in\theta},
\end{equation}
then we obtain the system of ordinary differential equations
\begin{equation}
  n{{dG_n}\over{dk}}+\tilde\nu k^2\left({{b+k^2}\over{1+k^2}}\right)G_n-
  {{ka}\over{2}}\left(G_{n+1}-G_{n-1}\right)=
  \left(\delta_{n,1}-\delta_{n,-1}\right)\delta(k).
  \label{gn}
\end{equation}
Note that the symmetry property
\begin{equation}
  G_n(-k)=-\left[G_{-n}(k)\right]^*
  \label{symmetry}
\end{equation}
follows from the fact that the enthalpy perturbation $h'$ is real.  As
a boundary condition, $G_n(k)$ is required to tend to zero as
$|k|\to\infty$, in order that it be the Fourier transform of a
continuous function.

The total tidal torque exerted on the disk is
\begin{equation}
  T=-\int_{-\infty}^\infty\int_0^{2\pi}\int_0^\infty
  (\rho+\rho')\partial_\phi\Phi'\,r\,dr\,d\phi\,dz.
\end{equation}
When expressed in terms of the scaled variables introduced in
Section~\ref{Reduction to dimensionless form}, the torque exerted in
the corotation region is
\begin{equation}
  T_{\rm c}={{mrc^2\Psi}\over{4\Omega}}
  {{d}\over{dr}}\left({{\Sigma}\over{B}}\right)
  \int_0^{2\pi}\int_{-\infty}^\infty f(x,\theta)\sin\theta\,dx\,d\theta,
\end{equation}
where the prefactor is understood to be evaluated at $r=r_{\rm c}$.
In terms of the Fourier representation, this further simplifies to
\begin{equation}
  T_{\rm c}=t_{\rm c}\,T_{\rm GT},
\end{equation}
where
\begin{equation}
  t_{\rm c}=G_1(0)-G_{-1}(0)
  \label{tc}
\end{equation}
is dimensionless, and
\begin{equation}
  T_{\rm GT}={{m\pi^2\Psi^2}\over{2(d\Omega/dr)}}
  {d\over{dr}}\left({{\Sigma}\over{B}}\right)
  \label{tgt}
\end{equation}
is the torque formula given by GT79.  Although equation (\ref{gn})
requires $G_{\pm1}(k)$ to have unit discontinuities at $k=0$, the
dimensionless torque $t_{\rm c}$ is nevertheless well defined as the
quantity $G_1(k)-G_{-1}(k)$ is continuous at $k=0$.

\subsection{A simplifying assumption}

\label{A simplifying assumption}

A considerable simplification in equation (\ref{gn}) is obtained when
$b=1$.  This occurs in a Keplerian disk when conditions in the
corotation region are such that $\nu\Sigma\propto\Sigma^{1/3}$
locally.  In the absence of any detailed knowledge of the effective
viscosity, we adopt this convenient assumption, anticipating that our
results will not depend sensitively on it.

It is then natural to rescale the wavenumber to
\begin{equation}
  \tilde k=\tilde\nu^{1/3}k,
\end{equation}
so that
\begin{equation}
  n{{dG_n}\over{d\tilde k}}+\tilde k^2G_n-
  p\tilde k\left(G_{n+1}-G_{n-1}\right)=
  \left(\delta_{n,1}-\delta_{n,-1}\right)\delta(\tilde k),
  \label{rescaled}
\end{equation}
where
\begin{equation}
  p={{1}\over{2}}a\tilde\nu^{-2/3}
\end{equation}
is the only remaining parameter of the problem.  In physical terms,
\begin{equation}
  p=\left({{\Psi}\over{-\kappa^2\,d\ln\Omega/d\ln r}}\right)
  \left({{-m\,d\Omega/dr}\over{\nu}}\right)^{2/3}
  \label{p}
\end{equation}
measures the strength of the potential relative to the viscosity
(i.e., the nonlinearity relative to the dissipation).  In terms of the
three characteristic length-scales defined in equation (\ref{delta}),
$p=(\delta_\Psi/\delta_\nu)^2$.

We may assume without loss of generality that $p>0$.  Our aim now is
to evaluate the function $t_{\rm c}(p)$ that determines how the
formula of GT79 is modified.

\subsection{Linear theory}

In linear theory the coupling parameter $p$ is set to zero and only
modes $n=\pm1$ are excited.  The solution is
\begin{equation}
  G_{\pm1}(\tilde k)=\pm H(\pm\tilde k)
  \exp(\mp{\textstyle{{1}\over{3}}}\tilde k^3),
  \label{g-lo-p}
\end{equation}
where $H$ is the Heaviside function.  The dimensionless torque
(eq.~[\ref{tc}]) is then $t_{\rm c}=1$.  This shows that the torque
formula of GT79, which was derived for a two-dimensional, inviscid
disk, also applies to a three-dimensional, barotropic and viscous
disk, provided that the linearization is valid (i.e., $p\ll1$).

\subsection{Feedback}

The axisymmetric component $G_0$ of the solution has a special role.
The case $n=0$ in equation (\ref{rescaled}) is not a differential
equation but states that
\begin{equation}
  \tilde k^2G_0=p\tilde k(G_1-G_{-1}).
  \label{k2g0}
\end{equation}
Now $G_0$ also appears in the equations for $G_{\pm1}$ in the form
$p\tilde kG_0$.  We solve equation (\ref{k2g0}) within the space of
generalized functions to obtain
\begin{equation}
  \tilde kG_0=p(G_1-G_{-1})+C_1\delta(\tilde k),
  \label{kg0}
\end{equation}
where $C_1$ is an arbitrary constant.  Equation (\ref{rescaled}) with
$n=\pm1$ then implies
\begin{equation}
  {{dG_1}\over{d\tilde k}}+\tilde k^2G_1-
  p\tilde kG_2+p^2(G_1-G_{-1})=(1-C_1p)\delta(\tilde k),
  \label{dg1}
\end{equation}
\begin{equation}
  {{dG_{-1}}\over{d\tilde k}}-\tilde k^2G_{-1}-
  p\tilde kG_{-2}+p^2(G_1-G_{-1})=(1-C_1p)\delta(\tilde k).
  \label{dg-1}
\end{equation}
The solution of equation (\ref{kg0}) is of the form
\begin{equation}
  G_0=G_0^{\rm reg}(\tilde k)+{p{t_{\rm c}}\over{\tilde k}}-
  C_1\delta'(\tilde k)+iC_2\delta(\tilde k),
  \label{g0}
\end{equation}
where $G_0^{\rm reg}(\tilde k)$ is regular at $\tilde k=0$, and $C_2$
is a second arbitrary constant.  Consider the inverse Fourier
transforms
\begin{equation}
  \int_{-\infty}^\infty k^{-1}\,e^{ikx}\,dk=i\pi\,{\rm sgn}(x),\qquad
  -\int_{-\infty}^\infty\delta'(k)\,e^{ikx}\,dk=ix,\qquad
  \int_{-\infty}^\infty i\,\delta(k)\,e^{ikx}\,dk=i,
\end{equation}
the first of which is a principal value.  By allowing $C_1\ne0$ we
would admit a perturbation that redefined the basic vortensity
gradient of the disk across the whole corotation region.  This should
be excluded, as discussed in Section~\ref{Reduction to dimensionless
  form}.  We may also set $C_2=0$ without loss of generality, as it
simply adds a constant to the basic surface density.

The solution generated by the $\tilde k^{-1}$ singularity in $G_0$
corresponds to a gradual, axisymmetric step in the enthalpy
perturbation across the corotation region, in the sense that
$f(x,\theta)$ tends to well defined limits as $x\to\pm\infty$, but
these limiting values are unequal.  This occurs because the tidal
torque causes a redistribution of the surface density of the disk.
Consider the standard evolutionary equations for an accretion disk in
which the corotation torque is added as an infinitely localized source,
\begin{equation}
  {{\partial\Sigma}\over{\partial t}}+
  {{1}\over{r}}{{\partial}\over{\partial r}}(r\Sigma v_r)=0,
\end{equation}
\begin{equation}
  \Sigma v_r{d\over{dr}}(r^2\Omega)=
  {{1}\over{r}}{{\partial}\over{\partial r}}
  \left(\nu\Sigma r^3{{d\Omega}\over{dr}}\right)+
  {{T_{\rm c}}\over{2\pi r_{\rm c}}}\,\delta(r-r_{\rm c}),
\end{equation}
where $v_r$ is the mean radial velocity.  In a steady state these
integrate to
\begin{equation}
  -2\pi r\Sigma v_r=\dot M={\rm constant},
\end{equation}
\begin{equation}
  \dot M r^2\Omega+2\pi\nu\Sigma r^3{{d\Omega}\over{dr}}+
  T_{\rm c}H(r-r_{\rm c})={\rm constant}.
\end{equation}
The step in $\nu\Sigma$ across the resonant region associated with the
$\tilde k^{-1}$ singularity in $G_0$ can be shown exactly to balance
the corotation torque $T_{\rm c}$ in this equation.  Note that this
simplified description does not treat the internal structure of the
corotation region, but is intended to show how the torque affects the
disk on a larger scale.

\subsection{Approximate solutions for small and large $p$}

\label{Approximate solutions}

Approximate solutions can be obtained by neglecting $G_2$ and $G_{-2}$
in equations (\ref{dg1}) and (\ref{dg-1}).  This amounts to a severe
Fourier truncation of the problem, in which we allow for the solution
to feed back on the axisymmetric part of the disk but neglect the
excitation of higher harmonics.  Nevertheless, this method is found to
give results in close agreement with the more accurate numerical
solutions described in the next section.

Defining $G_\pm=G_1\pm G_{-1}$ and taking the sum and difference of
equations (\ref{dg1}) and (\ref{dg-1}), we obtain
\begin{equation}
  {{dG_+}\over{d\tilde k}}+(\tilde k^2+2p^2)G_-=2\delta(\tilde k),
\end{equation}
\begin{equation}
  {{dG_-}\over{d\tilde k}}+\tilde k^2G_+=0,
\end{equation}
and so
\begin{equation}
  -{{d}\over{d\tilde k}}\left({{1}\over{\tilde k^2}}
  {{dG_-}\over{d\tilde k}}\right)+(\tilde k^2+2p^2)G_-=2\delta(\tilde k).
  \label{g-}
\end{equation}

For $p^2\ll1$ the solution may be expanded in powers of $p^2$,
\begin{equation}
  G_-=G_-^{(0)}+p^2G_-^{(1)}+O(p^4).
\end{equation}
Using the fact that the bounded solution of the equation
\begin{equation}
  -{{d^2y}\over{dx^2}}+y=2f(x)
\end{equation}
is
\begin{equation}
  y(x)=\int_{-\infty}^\infty f(x')\,e^{-|x-x'|}\,dx',
\end{equation}
we find, at successive orders in $p^2$,
\begin{equation}
  G_-^{(0)}(\tilde k)=\exp(-{\textstyle{{1}\over{3}}}|\tilde k^3|),
\end{equation}
\begin{equation}
  G_-^{(1)}(\tilde k)=-\int_{-\infty}^{\infty}
  \exp(-{\textstyle{{1}\over{3}}}|\tilde k'^3|-
  {\textstyle{{1}\over{3}}}|\tilde k^3-\tilde k'^3|)\,d\tilde k'.
\end{equation}
The dimensionless torque is then
\begin{equation}
  t_{\rm c}=G_-(0)=1-\left({{2}\over{3}}\right)^{2/3}
  \Gamma\left({{1}\over{3}}\right)p^2+O(p^4)
  \approx1-2.044\,p^2.
  \label{lo-approx}
\end{equation}
This asymptotic form can be shown to be correct even when the coupling
to higher harmonics is taken into account.  In this limit of small
$p$, only wavenumbers $|\tilde k|\la1$ contribute significantly to the
solution.

For $p^2\gg1$ we may neglect $\tilde k^2$ relative to $2p^2$ in
equation (\ref{g-}).  The solution is then
\begin{equation}
  G_-(\tilde k)\approx{{2^{1/8}}\over{\Gamma\left({{1}\over{4}}\right)}}
  p^{-3/4}|\tilde k|^{3/2}K_{3/4}\left(2^{-1/2}p\tilde k^2\right),
\end{equation}
where $K$ is the modified Bessel function of the second kind, and the
dimensionless torque is
\begin{equation}
  t_{\rm c}=G_-(0)\approx{{\Gamma\left({{3}\over{4}}\right)}\over
  {\Gamma\left({{1}\over{4}}\right)}}2^{1/4}p^{-3/2}\approx0.4019\,p^{-3/2}.
  \label{hi-approx}
\end{equation}
Since $p\propto\nu^{-2/3}$, this implies that the torque is
proportional to the viscosity when the viscosity is small.  This is to
be expected on general grounds, but to neglect the coupling to higher
harmonics is difficult to justify formally when $p$ is large.  The
above form of $G_{-}$ indicates that only wavenumbers $|\tilde k|\la
p^{-1/2}$ contribute significantly.

\subsection{Numerical solution}

\label{Numerical solution}

For a numerical solution, we truncate the system at some order $N$ and
set $G_n$ to zero for $|n|>N$.  We then solve equations (\ref{dg1})
and (\ref{dg-1}) together with equation (\ref{rescaled}) for
$2\le|n|\le N$.  Now the symmetry property (\ref{symmetry}) allows us
to consider $\tilde k>0$ only, and the jump conditions at $\tilde k=0$
imply
\begin{equation}
  G_n(0+)+G_{-n}(0+)=\delta_{n,1},\qquad n>0.
  \label{bc}
\end{equation}
We integrate the equations from $\tilde k=0+$ to a finite value
$\tilde k=\tilde k_{\rm max}$.  A `particular solution' $G_n^{\rm(p)}$
satisfying the inhomogeneous boundary condition (\ref{bc}) is
generated by starting from the initial condition $G_1=1$ ($G_n=0$
otherwise), and $N$ `complementary functions' $G_n^{({\rm c},q)}$ by
starting from initial conditions $G_q=-G_{-q}=1$ ($G_n=0$ otherwise)
for each $q=1,\dots,N$.  The amplitudes of the complementary functions
appearing in the desired solution are determined by requiring that
$G_n(\tilde k_{\rm max})=0$ for $-N\le n\le-1$.  This simulates the
requirement that $G_n$ should tend to zero as $|\tilde k|\to\infty$.
For $n<0$, $G_n$ has a tendency to grow as
$\exp({\textstyle{{1}\over{3}}}\tilde k^3/|n|)$ as $\tilde
k\to\infty$, and our boundary condition is designed to eliminate such
growing solutions.  For $n>0$, $G_n$ tends to decay naturally as
$\tilde k\to\infty$.

In Figure~1 we plot the dimensionless torque $t_{\rm c}$ determined
from the numerical solution as a function of the coupling parameter
$p$.  The solution converges rapidly as $N$ and $\tilde k_{\rm max}$
are increased, and good agreement is found with the approximate
solutions (\ref{lo-approx}) and (\ref{hi-approx}) for small and large
$p$.

\subsection{Interpretation}

We now attempt to give a real-space interpretation of these results.
Multiplying our governing equation (\ref{final}) by $-\Sigma
r/(2\Omega)$ and integrating over $\phi$, we obtain
\begin{equation}
  \partial_\xi^2\left\{-\Sigma\int\left[{{1}\over{\kappa^2}}
  (\partial_\phi\Phi')\partial_\xi h'+{{\nu r}\over{2\Omega}}
  \left(\partial_\xi^2+2r\Omega{{d\Omega}\over{dr}}{{D_\mu}
  \over{c^2}}\right)h'\right]\right\}\,d\phi=
  -{{\Sigma}\over{c^2}}\int(\partial_\phi\Phi')
  \partial_\xi h'\,d\phi,
\end{equation}
where the very first term has been taken over to the right-hand side.
The first integral with respect to $\xi$ is
\begin{equation}
  \partial_\xi\left\{-\Sigma\int\left[{{1}\over{\kappa^2}}
  (\partial_\phi\Phi')\partial_\xi h'+{{\nu r}\over{2\Omega}}
  \left(\partial_\xi^2+2r\Omega{{d\Omega}\over{dr}}
  {{D_\mu}\over{c^2}}\right)h'\right]\right\}\,d\phi=
  -{{\Sigma}\over{c^2}}\int h'\partial_\phi\Phi'\,d\phi,
  \label{am}
\end{equation}
plus a possible constant of integration.

Equation (\ref{am}) expresses the conservation of angular momentum in
a steady state, although the terms that balance in the unperturbed
disk do not appear.  The left-hand side is the divergence of the
angular momentum flux, and the right-hand side is the tidal torque per
unit radius.  To see this, the first term in the flux can be
identified as the integrated Reynolds stress
\begin{equation}
  r\int\!\!\int\rho u'v'\,d\phi\,dz, 
\end{equation}
using equations (\ref{eq2}) and (\ref{eq3}).  The remaining terms in
the flux are viscous stresses.  The right-hand side of equation
(\ref{am}) can also be written as
\begin{equation}
  -\int\!\!\int\rho'\partial_\phi\Phi'\,d\phi\,dz,
\end{equation}
which is the tidal torque per unit radius.

In the Fourier representation, when suitably normalized, equation
(\ref{am}) reads
\begin{equation}
  {{k^2a}\over{2}}(F_1-F_{-1})-\tilde\nu k(k^2+b)F_0=
  -{{a}\over{2}}(F_1-F_{-1}),
\end{equation}
plus a possible multiple of $\delta(k)$.  The same equation could be
written for $G$ instead of $F$, and is then equivalent to our equation
(\ref{gn}) in the case $n=0$.  (The possible additional term
proportional to $\delta(k)$ is equivalent to the $C_1$ term that
arises in passing from equation (\ref{k2g0}) to (\ref{kg0}), and can
be excluded on the same grounds.)  This shows how equation (\ref{gn})
partitions into terms associated with Reynolds stress, viscous torque
and tidal torque.

In Section~\ref{Approximate solutions} we worked with the quantity
$G_-\propto(1+k^2)(F_1-F_{-1})$.  In these terms, the Fourier
transforms of the Reynolds, viscous and tidal terms are proportional
to
\begin{equation}
  \left({{k^2}\over{1+k^2}}\right)G_-(k),\qquad
  G_-(k),\qquad
  \left({{1}\over{1+k^2}}\right)G_-(k),
\end{equation}
respectively.  When $p\ll1$ we found that $G_-(k)$ is cut off for
wavenumbers $|\tilde k|\ga1$, i.e. physical wavenumbers
$|k/\delta_c|\ga1/\delta_\nu$.  It follows that the characteristic scale
of the viscous torque is the viscous scale $\delta_\nu$ defined in
equation (\ref{delta}).  If the viscosity is small, so that
$\delta_\nu\ll\delta_c$, then the scale of the Reynolds stress is also
$\delta_\nu$.  However, the scale of the tidal torque is $\delta_c$,
because its Fourier transform is cut off for physical wavenumbers
$|k/\delta_c|\ga1/\delta_c$ by the factor $(1+k^2)^{-1}$.

When $p\gg1$ we found that $G_-(k)$ is cut off for wavenumbers
$|\tilde k|\ga p^{-1/2}$, i.e. physical wavenumbers
$|k/\delta_c|\ga1/\delta_\Psi$.  If the forcing amplitude is small, so
that $\delta_\Psi\ll\delta_c$, then the scale of the Reynolds stress is
also $\delta_\Psi$, but the scale of the tidal torque is again $\delta_c$.

To summarize, if $\delta_c$ is the largest of the three scales, then
the tidal torque is spread over a region of scale $\delta_c$, while
the Reynolds and viscous stresses have a smaller scale
$\max(\delta_\nu,\delta_\Psi)$.  This is consistent with the impression
obtained from the original theory (GT79), that the tidal torque
creates a disturbance in the disk on the scale $\delta_c$, which
subsequently transfers its angular momentum to the background disk,
through dissipation or nonlinearity, on a finer scale.

\section{Application to eccentric resonances}

\label{Application to eccentric resonances}

The reduction of the torque exchanged between the perturber and the
disk has consequences for the eccentricity evolution of young planets
orbiting in a protoplanetary disk.  We consider the first-order
eccentric corotation resonances associated with a planet of mass ratio
$q=M_{\rm p}/M_*$ executing a orbit of eccentricity $e$ within a
Keplerian disk.

GT80 provide expressions for the locations of these resonances and the
forcing potentials.  Inner and outer resonances occur at radii
\begin{equation}
  r=\left({{m}\over{m\pm 1}}\right)^{2/3}a_{\rm p},
\end{equation}
where $a_{\rm p}$ is the semi-major axis of the planet's orbit.  The
amplitude of the forcing potential may be written in the form
\begin{equation}
  {{\Psi}\over{r^2\Omega^2}}=0.8020\,C_m^\pm meq,
  \label{psi}
\end{equation}
where $C_m^\pm$ is a correction factor that tends to unity for
large $m$.  Accordingly, the saturation parameter of our analysis is
\begin{equation}
  p=0.7006\,C_m^\pm m^{5/3}eq\alpha^{-2/3}
  \left({{H}\over{r}}\right)^{-4/3},
\end{equation}
where we write $\nu=\alpha cH$ and $H=c/\Omega$.  When $p=1$, i.e., when
\begin{equation}
  e=e_{0.63}={{1.427}\over{C_m^\pm}}m^{-5/3}q^{-1}\alpha^{2/3}
  \left({{H}\over{r}}\right)^{4/3},
  \label{e_063}
\end{equation}
$63\%$ saturation occurs because $t_{\rm c}\approx0.37$.  For $5\%$
saturation, only $p\approx0.16$ is required.

Table~1 contains results for the lowest-order outer eccentric
resonances ($l=m-1$ in the notation of GT80) for a disk with
$H/r=0.05$ and a planet of mass ratio $q=0.001$.  The first column
labels the azimuthal wavenumber of the potential component being
considered.  The second column gives the location of the resonance in
units of $a_{\rm p}$.  The third column contains the correction factor
defined in equation (\ref{psi}).  The fourth column gives a measure of
the relative importance of the resonances: neglecting the torque
cut-off and any differences in the disk parameters between different
resonant radii, the unsaturated torque scales $\propto m\Psi^2\propto
C_m^2m^3$.  The fifth column, labeled $e_{0.63}$, is the eccentricity
required to reduce the eccentric corotation torque $63\%$ below its
unsaturated value for a turbulent viscosity parameter $\alpha=0.004$.
The sixth column, $e_{0.05}$, is similar to the fifth column, but
gives the eccentricity needed to reduce the torque $5\%$ below its
unsaturated level.  The final column, $\alpha_{0.63}$, is the value of
$\alpha$ corresponding to $63\%$ saturation when the eccentricity is
$e=0.1$.  Table~2 gives the same data for the inner resonances
($l=m+1$).

Resonances of high order occur very close to the planet's orbit, and
are present only if the planet does not clear a substantial gap in the
disk.  An analysis of such resonances requires a consideration of a
number of effects that we have neglected.  When $m$ becomes comparable
to $r/H$, additional terms in the dynamical equations must be
included, and the torque is reduced from the standard value (GT79).
The proximity of the resonance to the planet also means that the
perturbing potential varies significantly with $z$ within the disk,
and the relevant quantity is the density-weighted vertical average of
the potential, rather than the mid-plane value (Ward 1986, 1989).
These two effects apply equally to Lindblad and corotation resonances,
and so do not alter the $5\%$ balance between torques that excite and
damp eccentricity.  A further effect is that neighboring resonances of
sufficiently high order overlap one another, which may alter the
resonant torques.

\section{Summary and discussion}

\label{Summary and discussion}

We have determined the torque exerted in a steady state by an external
potential on a three-dimensional gaseous disk at a non-coorbital
corotation resonance.  Our model accounts for the feedback of the
torque on the surface density and vorticity in the corotation region,
and assumes that the disk has a barotropic equation of state and a
nonzero effective viscosity.  The torque formula of Goldreich \&
Tremaine (1979) (eq.~[\ref{tgt}]) must be modified by a reduction
factor $t_{\rm c}(p)$, plotted in Figure~1, which quantifies the
extent to which the resonance is saturated.  This factor depends
essentially on a single dimensionless parameter $p$, defined in
equation (\ref{p}), which measures the strength of the potential
$\Psi$ relative to the viscosity $\nu$ (i.e., the nonlinearity
relative to the dissipation).

In Section~\ref{Numerical solution}, we determined that the
characteristic radial width of the resonance in the limit of large $p$
(low viscosity) is $\delta\sim\sqrt{\Psi}/\Omega$.  (This refers to
the scale of Reynolds and viscous stresses; the tidal torque is spread
over a region of characteristic width $c/\kappa$.)  We found that the
torque is reduced by a factor $\sim p^{-3/2}$, which can be regarded
as the ratio of the viscous diffusion rate $\sim \nu/\delta^2$ over
the libration region to the libration rate $\sim m|d\Omega/dr|\delta$.
In this regime, our results are broadly consistent with the saturation
model of Ward (1992), which involves the same ratio of rates, but
applied to the coorbital region.  Earlier, Goldreich \& Tremaine
(1981) had argued that, in a disk of collisional particles, saturation
would occur when (in our notation) $p\gg1$.

Goldreich \& Sari (2002) have recently developed an evolutionary model
for planetary eccentricity in which eccentricity growth occurs through
a finite-amplitude instability.  For infinitesimal eccentricity, the
damping caused by corotation resonances just overcomes the
eccentricity growth due to Lindblad resonances.  Above a critical
level of eccentricity, however, the corotation resonances become
sufficiently saturated that growth occurs.  Using our evaluation of
the saturation function $t_{\rm c}(p)$, Goldreich \& Sari (2002) were
able to determine the critical value of eccentricity required within
the context of a specific disk model.  The results in
Section~\ref{Application to eccentric resonances} and Table~1 show
that for a Jupiter-mass planet orbiting within a typical
protoplanetary disk, eccentricities of a few percent are adequate to
saturate all first-order eccentric corotation resonances except those
of the lowest $m$-values ($m\la4$).  To achieve a $5\%$ reduction in
torque at such resonances, as may be sufficient to change the balance
from eccentricity damping to growth (GT80), requires only
eccentricities of $1\%$ or less.

Exactly which resonances contribute most to eccentricity evolution
depends on the extent of the gap cleared by the planet, which in turn
depends on the mass ratio and the properties of the disk.  As seen in
Table~1, the largest corotation torques are those associated with the
highest $m$-values, and they are the easiest to saturate.  A
lower-mass planet that opens a smaller gap may excite many resonances
so that the dominant torque comes from approximately the cut-off
$m$-value, of order $r/H$ (GT80).  In that case, $63\%$ saturation is
achieved optimistically at the torque cut-off when
$e\ga1.4\,q^{-1}\alpha^{2/3}(H/r)^3$.  Note that, although the large
value of $m$ is advantageous for saturation, the low mass of the
planet is unfavorable (see eq.~[\ref{e_063}]).
On the other hand, if the planet mass is small enough so that there is
no gap in the disk, then the coorbital resonances cause eccentricity
decay (Ward 1988; Artymowicz 1993).

We have neglected a number of potential complications.  Numerical
simulations of a Jupiter-mass planet in a circular orbit with the disk
parameters adopted in Section~\ref{Application to eccentric
  resonances} suggest that the disk edge is not sharp (i.e., its
radial extent is much greater than $H$; see Figure~1 of Lubow,
Seibert, \& Artymowicz 1999).  It is evident, however, that within the
radial extent of the planet's Roche lobe, material is captured by the
planet and the resonant effects considered here do not play a role.
It is not clear which eccentric corotation resonances are excited in
the presence of an eccentric planet.  The complications are due at
least in part to nonlinear effects other than those considered here.
For example, shocks associated with the planet's wake cause a
non-closure of streamlines in the vicinity of the disk edge, leading
to a drift of material through the corotation regions.  In addition,
material within the Roche lobe of the planet exerts torques that may
contribute to the eccentricity balance.

As usual in accretion disk theory, our knowledge of the effective
viscosity of the disk is limited.  In Section~\ref{A simplifying
  assumption} we adopted a convenient assumption in order to make
analytical progress.  Although we have succeeded in analyzing a
three-dimensional disk, thereby generalizing existing theories, we
assumed that the disk was barotropic.  The effects of buoyancy or
baroclinicity on the corotation region remain uninvestigated.

In recent work we were able to verify analytical theories of the
torques exerted at Lindblad and vertical resonances through direct
numerical simulations (Bate et al. 2002).  It would be valuable to
conduct simulations of a non-coorbital corotation resonance to test
the findings of the present paper.

After eccentric corotation resonances saturate at small eccentricity,
the eccentricity growth may be limited at intermediate values.  The
limitation may be due to the overlap of resonances or alternatively
through the excitation of higher order Lindblad resonances, some of
which cause eccentricity damping.  A contribution to eccentricity
damping occurs at an eccentric inner (outer) Lindblad resonance that
lies outside (inside) the planet's orbit.  SPH simulations of
eccentric-orbit binary stars suggest that little eccentricity growth
via resonances occurs for eccentricities in the range of $0.5-0.7$ or
higher (Lubow \& Artymowicz 1993), and similar limits may occur for
planets.

In conclusion, we have found a simple quantitative measure of the
saturation of a corotation resonance in a gaseous disk.  This analysis
suggests that planets may plausibly experience a net growth of
eccentricity through their interaction with the disk in a variety of
circumstances, provided that the eccentricity is not extremely small
to begin with.

\acknowledgments

We are grateful to Peter Goldreich and Re'em Sari for allowing us to
coordinate our presentation of results with theirs prior to
publication, and also for correcting some inaccuracies and omissions
in an earlier version of this paper.  We also thank the anonymous
referee for a very thorough report that led to an improvement of the
paper.  This work was initiated at the Aspen Center for Physics, and
we are grateful for their hospitality.  GIO acknowledges the support
of the Royal Society through a University Research Fellowship.  SHL
acknowledges support from NASA grant NAG5-10732.

\newpage

\begin{deluxetable}{rrrrrrr}
  \tablewidth{12cm}
  \tablecaption{Saturation parameters for outer corotation resonances}
  \tablehead{\colhead{$m$}&\colhead{$r/a_{\rm p}$}&\colhead{$C_m^-$}&
    \colhead{$C_m^2m^3$}&\colhead{$e_{0.63}$}&\colhead{$e_{0.05}$}&
    \colhead{$\alpha_{0.63}$}}
  \startdata
    $2$&$1.5874$&$0.7422$&$4.407$&$0.2812$&$0.0453$&$0.00085$\\
    $3$&$1.3104$&$0.8418$&$19.13$&$0.1261$&$0.0203$&$0.00282$\\
    $4$&$1.2114$&$0.8854$&$50.18$&$0.0742$&$0.0120$&$0.00625$\\
    $5$&$1.1604$&$0.9101$&$103.5$&$0.0498$&$0.0080$&$0.01139$\\
    $6$&$1.1292$&$0.9261$&$185.2$&$0.0361$&$0.0058$&$0.01843$\\
    $7$&$1.1082$&$0.9372$&$301.3$&$0.0276$&$0.0044$&$0.02759$\\
    $8$&$1.0931$&$0.9454$&$457.6$&$0.0219$&$0.0035$&$0.03903$\\
    $9$&$1.0817$&$0.9517$&$660.3$&$0.0179$&$0.0029$&$0.05292$\\
   $10$&$1.0728$&$0.9567$&$915.3$&$0.0149$&$0.0024$&$0.06941$\\
  \enddata
\end{deluxetable}

\begin{deluxetable}{rrrrrrr}
  \tablewidth{12cm}
  \tablecaption{Saturation parameters for inner corotation resonances}
  \tablehead{\colhead{$m$}&\colhead{$r/a_{\rm p}$}&\colhead{$C_m^+$}&
    \colhead{$C_m^2m^3$}&\colhead{$e_{0.63}$}&\colhead{$e_{0.05}$}&
    \colhead{$\alpha_{0.63}$}}
  \startdata
    $1$&$0.6300$&$0.3365$&$0.113$&$1.9688$&$0.3170$&$0.00005$\\
    $2$&$0.7631$&$1.1819$&$11.17$&$0.1766$&$0.0284$&$0.00170$\\
    $3$&$0.8255$&$1.1265$&$34.26$&$0.0942$&$0.0152$&$0.00437$\\
    $4$&$0.8618$&$1.0970$&$77.02$&$0.0599$&$0.0096$&$0.00862$\\
    $5$&$0.8855$&$1.0787$&$145.5$&$0.0420$&$0.0068$&$0.01469$\\
    $6$&$0.9023$&$1.0663$&$245.6$&$0.0314$&$0.0050$&$0.02277$\\
    $7$&$0.9148$&$1.0572$&$383.4$&$0.0245$&$0.0039$&$0.03305$\\
    $8$&$0.9245$&$1.0503$&$564.8$&$0.0197$&$0.0032$&$0.04570$\\
    $9$&$0.9322$&$1.0449$&$795.9$&$0.0163$&$0.0026$&$0.06088$\\
   $10$&$0.9384$&$1.0406$&$1083.$&$0.0137$&$0.0022$&$0.07873$\\
  \enddata
\end{deluxetable}

\newpage

\begin{figure}
  \plotone{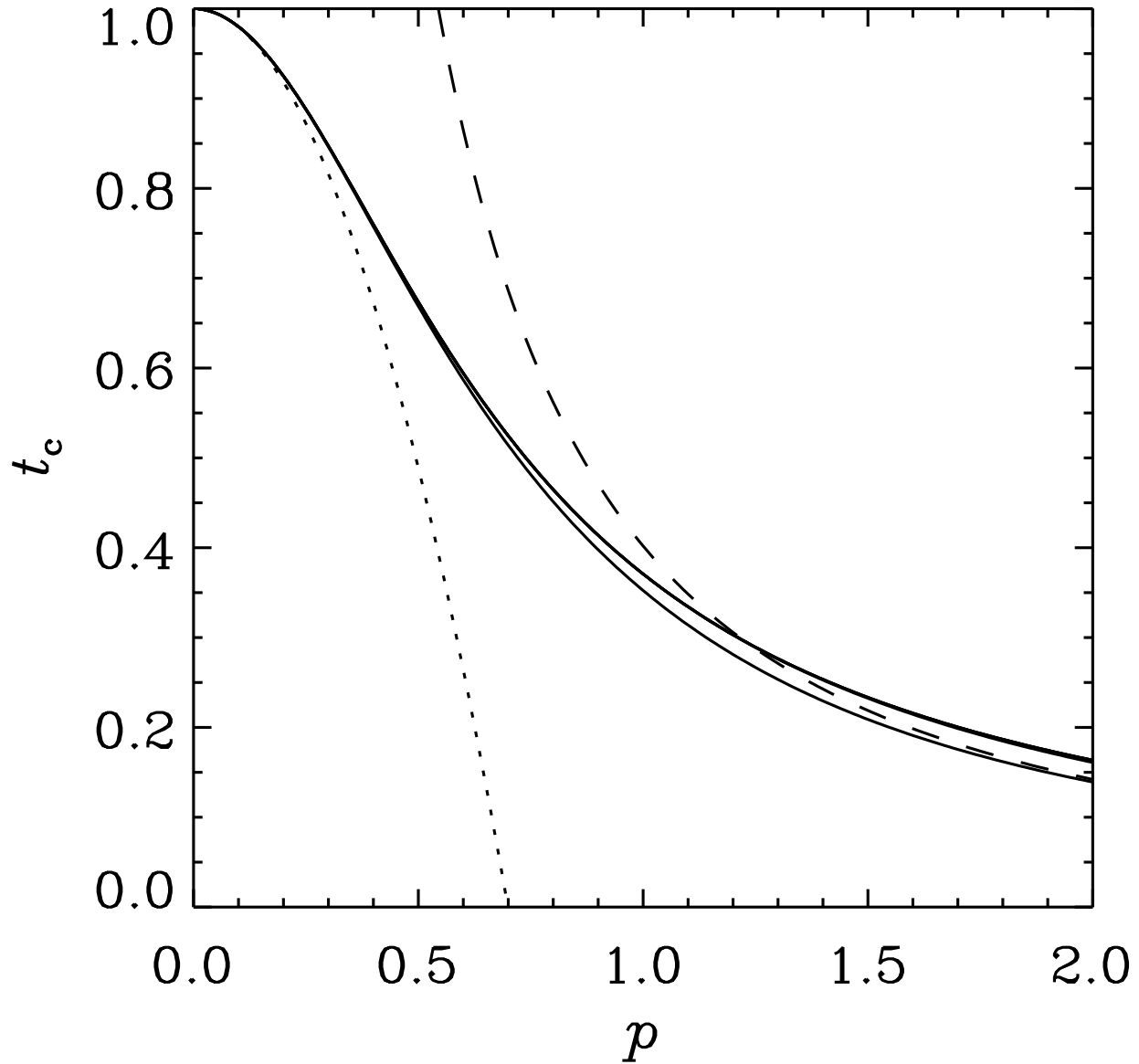}
  \figcaption{The dimensionless torque versus the coupling parameter.
    The curves calculated for truncation orders $N=1,\dots,5$ and
    $\tilde k_{\rm max}=3$ are plotted as solid lines; the curve for
    $N=1$ lies slightly below the others, while those for $N>1$ are
    indistinguishable by eye.  The dotted line shows the small-$p$
    approximation (\ref{lo-approx}) and the dashed line shows the
    large-$p$ approximation (\ref{hi-approx}), to which the curve for
    $N=1$ asymptotes.}
\end{figure}

\end{document}